\documentclass[prl,twocolumn,reprint,amsmath,amssymb,amsfonts,superscriptaddress]{revtex4-1}

\usepackage{graphicx}

\begin{document}

\title{Observation of dispersive shock waves, solitons, and their
  interactions in viscous fluid conduits}

\author{Michelle D.~\surname{Maiden}}
\affiliation{Department of Applied Mathematics, University of Colorado,
  Boulder CO 80309, USA}
\author{Nicholas K.~\surname{Lowman}}
\affiliation{Department of Mathematics, North
  Carolina State University, Raleigh, North Carolina 27695, USA}
\author{Dalton V.~\surname{Anderson}}
\author{Marika E.~\surname{Schubert}}
\author{Mark A.~\surname{Hoefer}}
\email{hoefer@colorado.edu}
\affiliation{Department of Applied Mathematics, University of Colorado,
  Boulder CO 80309, USA}
\date{\today}

\begin{abstract}
  Dispersive shock waves and solitons are fundamental nonlinear
  excitations in dispersive media, but dispersive shock wave studies
  to date have been severely constrained.  Here we report on a novel
  dispersive hydrodynamics testbed: the effectively frictionless
  dynamics of interfacial waves between two high contrast, miscible,
  low Reynolds' number Stokes fluids.  This scenario is realized by
  injecting from below a lighter, viscous fluid into a column filled
  with high viscosity fluid.  The injected fluid forms a deformable
  pipe whose diameter is proportional to the injection rate, enabling
  precise control over the generation of symmetric interfacial waves.
  Buoyancy drives nonlinear interfacial self-steepening while normal
  stresses give rise to dispersion of interfacial waves.  Extremely
  slow mass diffusion and mass conservation imply that the interfacial
  waves are effectively dissipationless.  This enables high fidelity
  observations of large amplitude dispersive shock waves in this
  spatially extended system, found to agree quantitatively with a
  nonlinear wave averaging theory.  Furthermore, several highly
  coherent phenomena are investigated including dispersive shock wave
  backflow, the refraction or absorption of solitons by dispersive
  shock waves, and the multi-phase merging of two dispersive shock
  waves.  The complex, coherent, nonlinear mixing of dispersive shock
  waves and solitons observed here are universal features of
  dissipationless, dispersive hydrodynamic flows.
\end{abstract}

\pacs{
  }

\maketitle

The behavior of a fluid-like, dispersive medium that exhibits
negligible dissipation is spectacularly realized during the process of
wave breaking that generates coherent nonlinear wavetrains called
dispersive shock waves (DSWs).  A DSW is an expanding, oscillatory
train of amplitude-ordered nonlinear waves composed of a large
amplitude solitonic wave adjacent to a monotonically decreasing wave
envelope that terminates with a packet of small amplitude dispersive
waves.  Thus, DSWs coherently encapsulate a range of fundamental,
universal features of nonlinear wave systems.  More broadly, DSWs
occur in dispersive hydrodynamic media that exhibit three unifying
features: i) nonlinear self-steepening, ii) wave dispersion, iii)
negligible dissipation (c.f.~the comprehensive DSW review article
\cite{el_dispersive_2016}).

Dispersive shock waves and solitons are ubiquitous excitations in
dispersive hydrodynamics, having been observed in many environments
such as quantum shocks in quantum systems (ultra-cold atoms
\cite{dutton_observation_2001,hoefer_dispersive_2006-1}, semiconductor
cavities \cite{amo_polariton_2011}, electron beams
\cite{mo_experimental_2013}), optical shocks in nonlinear photonics
\cite{rothenberg_observation_1989}, undular bores in geophysical
fluids \cite{hammack_korteweg-vries_1978,farmer_generation_1999}, and
collisionless shocks in rarefied plasma
\cite{taylor_observation_1970}.  However, all DSW studies to date have
been severely constrained by expensive laboratory setups
\cite{dutton_observation_2001,hoefer_dispersive_2006-1,mo_experimental_2013,hammack_korteweg-vries_1978}
or challenging field studies \cite{farmer_generation_1999},
difficulties in capturing dynamical information
\cite{rothenberg_observation_1989,dutton_observation_2001,hoefer_dispersive_2006-1},
complex physical modeling \cite{farmer_generation_1999}, or a loss of
coherence due to multi-dimensional instabilities
\cite{dutton_observation_2001,amo_polariton_2011} or dissipation
\cite{taylor_observation_1970,mo_experimental_2013}.  Here we report
on a novel dispersive hydrodynamics testbed that circumvents all of
these difficulties: the effective superflow of interfacial waves
between two high viscosity contrast, low Reynolds number Stokes
fluids.  The viscous fluid conduit system was well-studied in the
1980s as a simplified model of magma transport through the Earth's
partially molten upper mantle
\cite{scott_magma_1984,scott_observations_1986,whitehead_wave_1988}
(see also the background material in \cite{_see_????}).  This system
enables high fidelity studies of large amplitude DSWs, which are found
to agree quantitatively with nonlinear wave averaging or Whitham
theory
\cite{whitham_linear_1974,gurevich_nonstationary_1974,lowman_dispersive_2013}.
We then report the first experimental observations of highly coherent
phenomena including DSW backflow, the refraction or absorption of
solitons interacting with DSWs, and multi-phase DSW-DSW merger.  In
addition to its fundamental interest, the nonlinear mixing of
mesoscopic scale solitons and macroscopic scale DSWs could play a
major role in the initiation of decoherence and a one-dimensional,
integrable turbulent state \cite{el_kinetic_2005} that has recently
been observed in optical fibers \cite{randoux_intermittency_2014} and
surface ocean waves \cite{costa_soliton_2014}.

\begin{figure*}
  \centering
  \includegraphics{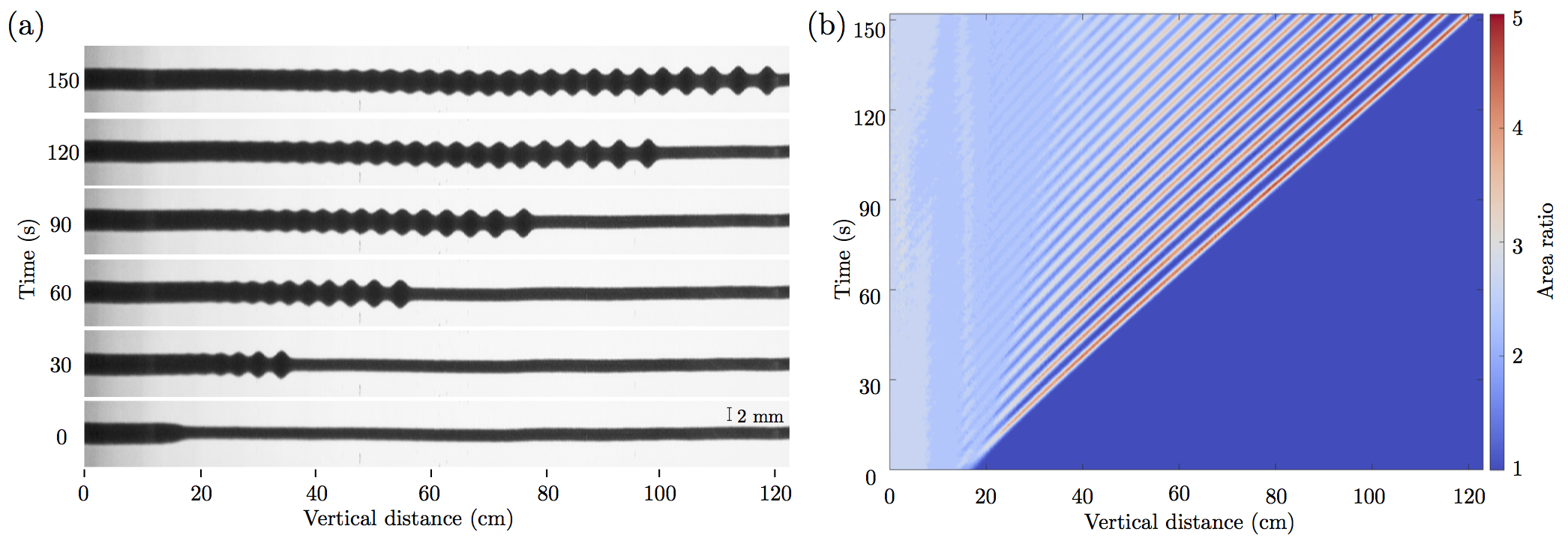}
  \caption{Interfacial wave breaking of two Stokes fluids causing the
    spontaneous emergence of coherent oscillations, a DSW.  The
    leading, downstream edge is approximately a large amplitude
    soliton whose phase speed is tied to the upstream conduit area.
    The trailing, upstream edge is a small amplitude wave packet
    moving at the group velocity whose wavenumber is tied to the
    downstream conduit area.  (a) 90$^\circ$ clockwise rotated,
    time-lapse digital images (aspect ratio 10:1).  (b) Space-time
    contour plot of the conduit cross-sectional area from (a). Nominal
    experimental parameters: $\Delta \rho = 0.0928$ g/cm$^3$,
    $\mu_{\rm i} = 91.7$ cP, $\epsilon = 0.030$, downstream flow rate
    $Q_0 = 0.50$ mL/min, and $a_- = 2.5$.}
  \label{fig:dsw}
\end{figure*}
In our experiment, the steady injection of an intrusive viscous fluid
(dyed, diluted corn syrup) into an exterior, miscible, much more
viscous fluid (pure corn syrup) leads to the formation of a stable
fluid filled pipe or conduit \cite{whitehead_1975}.  Due to high
viscosity contrast, there is minimal drag at the conduit interface so
the flow is well approximated by the Poiseulle or pipe flow relation
$D \propto Q^{1/4}$ where $Q$ is the injection rate and $D$ is the
conduit diameter.  By modulating the injection rate, interfacial wave
dynamics ensue.  Dilation of the conduit gives rise to buoyancy
induced nonlinear self-steepening regularized by normal interfacial
stresses that manifest as interfacial wave dispersion
\cite{olson_solitary_1986,lowman_dispersive_2013-1}.  Negligible mass
diffusion implies a sharp conduit interface and conservation of
injected fluid.  By identifying the azimuthally symmetric
\textit{conduit interface} as our one-dimensional dispersive
hydrodynamic medium, we arrive at the counterintuitive behavior that
viscous dominated, Stokes fluid dynamics exhibit dissipationless or
frictionless interfacial wave dynamics.  This will be made
mathematically precise below.  

By gradually increasing the injection rate, we are able to initiate
the spontaneous emergence of interfacial wave oscillations on an
otherwise smooth, slowly varying conduit.  See \cite{_see_????}  for
additional experimental details.  Figure \ref{fig:dsw}(a) displays a
typical time-lapse of our experiment.  At time $0$ s, the conduit
exhibits a relatively sharp transition between narrower and wider
regions.  Due to buoyancy, the interface of the wider region moves
faster than the narrower region.  Rather than experience folding over
on itself, the interface begins to oscillate due to dispersive effects
as shown in Fig.~\ref{fig:dsw}(a) at 30 s.  As later times in
Fig.~\ref{fig:dsw}(a) attest, the oscillatory region expands while the
oscillation amplitudes maintain a regular, rank ordering from large to
small.  By extracting the spatial variation of the normalized conduit
cross-sectional area $a$
from a one frame per second image sequence,
we display in Fig.~\ref{fig:dsw}(b) the full spatio-temporal
interfacial dynamics as a contour plot.  This plot
reveals two characteristic fronts associated with the oscillatory
dynamics: a large amplitude leading edge and a small amplitude,
oscillatory envelope trailing edge.

We can interpret these dynamics as a DSW resulting from the physical
realization of the Gurevich-Pitaevskii (GP) problem
\cite{gurevich_nonstationary_1974}, a standard textbook problem for
the study of DSWs \cite{el_dispersive_2016} that has been inaccessible
in other dispersive hydrodynamic systems.  Here, the GP problem is the
dispersive hydrodynamics of an initial jump in conduit area.  Although
we have only boundary control of the conduit width, our carefully
prescribed injection protocol \cite{_see_????} enables delayed
breaking far from the injection site.  This allows for the isolated
creation and long-time propagation of a ``pure'' DSW connecting two
uniform, distinct conduit areas.  Related excitations in the conduit
system were previously interpreted as periodic wave trains modeling
mantle magma transport \cite{scott_observations_1986}.  As we now
demonstrate, the interfacial dynamics observed here exhibit a
soliton-like leading edge propagating with a well-defined nonlinear
phase velocity, an interior described by a modulated nonlinear
traveling wave, and a harmonic wave trailing edge moving with the
linear group velocity.  The two distinct speeds of wave propagation in
one coherent structure are a striking realization of the double
characteristic splitting from linear wave theory
\cite{whitham_linear_1974}.

The long wavelength approximation of the interfacial fluid dynamics is
the conduit equation
\cite{scott_observations_1986,lowman_dispersive_2013-1}
\begin{equation}
  \label{eq:conduit}
  a_t + \left ( a^2 \right )_z - \left ( a^2
    \left(a^{-1} a_t \right)_z \right)_z  = 0 \ .
\end{equation}
Here, $a(z,t)$ is the nondimensional cross-sectional area of the
conduit as a function of the scaled vertical coordinate $z$ and time
$t$ (subscripts denote partial derivatives).  Both the interface of
the experimental conduit system and equation \eqref{eq:conduit}
exhibit the essential features of frictionless, dispersive
hydrodynamics: nonlinear self-steepening (second term) due to buoyant
advection of the intrusive fluid, dispersion (third term) from normal
stresses, and no dissipation due to the combination of intrusive fluid
mass conservation and negligible mass diffusion \cite{_see_????}.  The
analogy to frictionless flow corresponds to the \textit{interfacial}
dynamics, not the momentum diffusion dominated flow of the bulk.  The
conduit equation \eqref{eq:conduit} is nondimensionalized according to
cross-sectional area, vertical distance, and time in units of $A_0 =
\pi R_0^2$, $L_0 = R_0/\sqrt{8 \epsilon}$, and $T_0 = \mu_{\rm i}/L_0
g \Delta\rho \epsilon$, respectively, where $R_0$ is the downstream
conduit radius, $\epsilon = \mu_{\rm i}/\mu_{\rm e}$ is the viscosity
ratio of the intrusive to exterior liquids, $\Delta \rho = \rho_{\rm
  e} - \rho_{\rm i}$ is the density difference, and $g$ is gravity
acceleration.  Initially proposed as a simplified model for the
vertical ascent of magma along narrow, viscously deformable dikes and
principally used to study solitons
\cite{scott_observations_1986,olson_solitary_1986,helfrich_solitary_1990},
the conduit equation \eqref{eq:conduit} has since been derived
systematically from the full set of coupled Navier-Stokes fluid
equations using a perturbative procedure with the viscosity ratio as
the small parameter \cite{lowman_dispersive_2013-1}.  The conduit
equation \eqref{eq:conduit} was theoretically shown to be valid for
long times and large amplitudes under modest physical assumptions on
the basin geometry, background velocities, fluid compositions, weak
mass to momentum diffusion, and characteristic aspect ratio.  The
efficacy of this model has been experimentally verified in the case of
solitons
\cite{olson_solitary_1986,helfrich_solitary_1990}.

\begin{figure}
  \centering
  \includegraphics{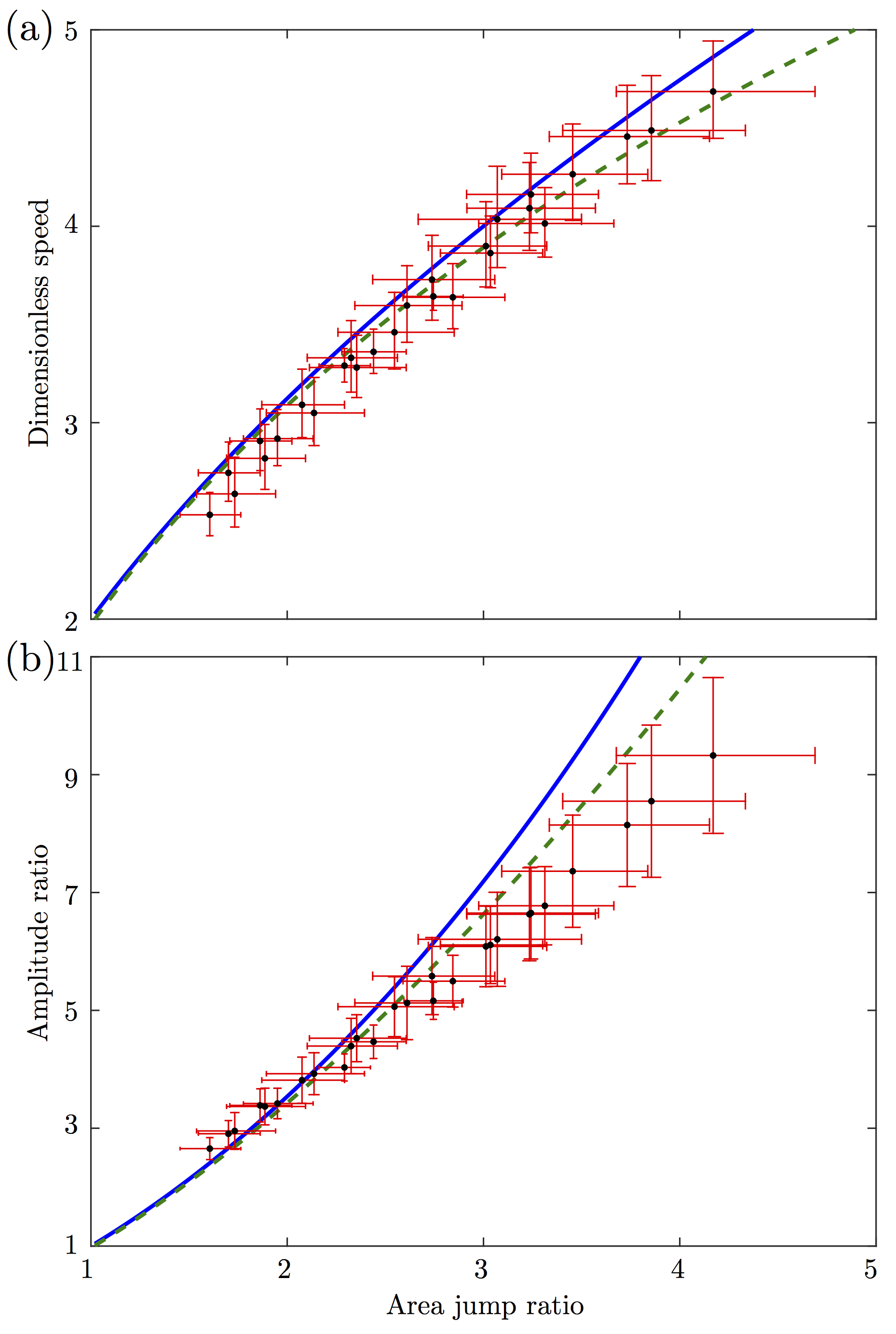}
  \caption{Comparison of observed and predicted leading edge DSW
    amplitude and speed.  Observations (circles), Whitham modulation
    theory (solid), and numerical simulation of the conduit equation
    (dashed) for (a) DSW leading edge speeds $s_+$ and (b) DSW leading
    amplitude $a_+$ versus downstream area ratio $a_-$.  Nominal
    experimental parameters: $\Delta \rho = 0.1305$ g/cm$^3$,
    $\mu_{\rm i} = 80.4$ cP (measured), $\mu_{\rm i} = 104$ cP
    (fitted), $\epsilon = 0.0024$.  See \cite{_see_????}  for fitting
    procedure.}
  \label{fig:expt}
\end{figure}
The study of DSWs involves a nonlinear wave modulation theory,
commonly referred to as Whitham theory \cite{whitham_linear_1974},
which treats a DSW as an adiabatically modulated periodic wave
\cite{gurevich_nonstationary_1974,el_dispersive_2016}.  Using Whitham
theory and eq.~\eqref{eq:conduit}, key conduit DSW physical features
such as leading soliton amplitude and leading/trailing speeds have
been determined \cite{lowman_dispersive_2013}.  For the jump in
downstream to upstream area ratio $a_-$, Whitham theory applied to the
conduit equation \eqref{eq:conduit} predicts relatively simple
expressions for the DSW leading $s_+$ and trailing $s_-$ edge speeds
\begin{equation}
  \label{eq:1}
  s_+ = \sqrt{1 + 8 a_-}-1, \quad s_- = 3 + 3 a_- - 3\sqrt{a_-(8 +
    a_-)} ,
\end{equation}
in units of the characteristic speed $U_0 = L_0/T_0$.  The leading
edge approximately corresponds to an isolated soliton where the
modulated periodic wave exhibits a zero wavenumber.  Given the speed
$s_+$, the soliton amplitude $a_+$ is implicitly determined from the
soliton dispersion relation $s_+ = [a_+^2(2 \ln a_+ - 1) +
1]/(a_+-1)^2$ \cite{olson_solitary_1986}.  At the trailing edge, the
modulated wave limits to zero amplitude, corresponding to harmonic
waves propagating with the group velocity $s_- = \omega'(k_-)$, where
$\omega(k) = 2a_-k/(1 + a_-k^2)$ is the linear dispersion relation of
eq.~\eqref{eq:conduit} on a background conduit area $a_-$ and $k_-^2 =
(a_- - 4 + \sqrt{a_-(8+a_-)})/(4a_-)$ is the distinguished wavenumber
determined from modulation theory \cite{lowman_dispersive_2013} (see
also \cite{el_dispersive_2016}).


In Fig.~\ref{fig:expt}, we compare the leading edge amplitude and
speed predictions with experiment, demonstrating quantitative
agreement for a range of jump values $a_-$.  The analytical theory
(Whitham theory) is known to break down at large amplitudes
\cite{lowman_dispersive_2013} so we also include direct determination
of the speed and amplitude from numerical simulation of
eq.~\eqref{eq:conduit}, demonstrating even better agreement.  In order
to obtain the reported dimensionless speeds of Fig.~\ref{fig:expt}(a),
we divide the measured speeds by $U_0$ with $\mu_{\rm i}$ determined
by fitting the downstream conduit area to a Poiseulle flow relation.
This enables us to self-consistently account for the shear-thinning
properties of corn syrup.  All the remaining fluid parameters take
their nominal, measured values.  The deviation between experiment and
theory at large jump values is consistent with previous measurements
of solitons, where the soliton dispersion relation was found to
underpredict observed speeds at large amplitudes
\cite{olson_solitary_1986} (see also \cite{_see_????}).

\begin{figure}
  \centering
  \includegraphics{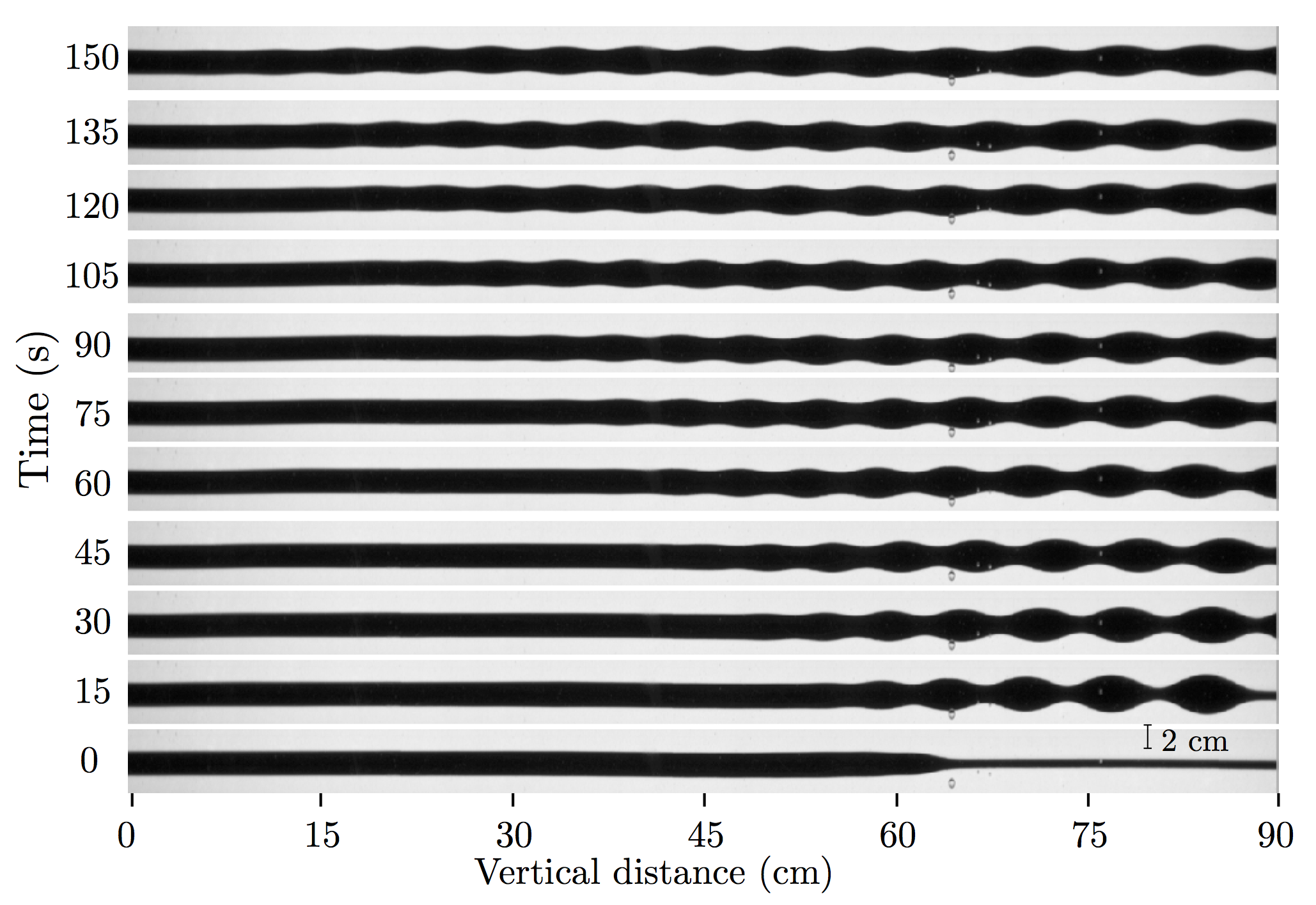}
  \caption{Time-lapse images (aspect ratio 1:1) of large amplitude
    wave breaking leading to upstream propagation of the DSW trailing
    edge envelope: DSW backflow.  Nominal experimental parameters:
    $\Delta \rho = 0.0983$ g/cm$^3$, $\mu_{\rm i} = 93.5$ cP, $\epsilon
    = 0.029$, $a_- = 4$, and $Q_0 = 0.50$ mL/min.}
  \label{fig:backflow}
\end{figure}
In addition to single DSWs, our experimental setup allows us to
investigate exotic, coherent effects predicted by
eq.~\eqref{eq:conduit} for the first time.  For example, backflow is a
feature of dispersive hydrodynamic systems whereby a portion of the
DSW envelope propagates upstream.  This feature occurs here when the
group velocity of the trailing edge wave packet is negative.  From the
expression for $s_-$ in \eqref{eq:1}, we predict the onset of backflow
when $a_-$ exceeds $8/3$.  In Fig.~\ref{fig:backflow}, we utilize our
injection protocol to report the observation of this phenomenon in the
viscous conduit setting (see \cite{_see_????} for video).  Waves with
strictly positive phase velocity are continually generated at the
trailing edge but the envelope group velocity is negative.
We estimate the crossover to backflow for the experiments reported in
Fig.~\ref{fig:expt} at $a_- \approx 3$, consistent with a slightly
larger crossover than theory (8/3) due to sub-imaging-resolution of
small amplitude waves.

\begin{figure*}
  \centering
  \includegraphics{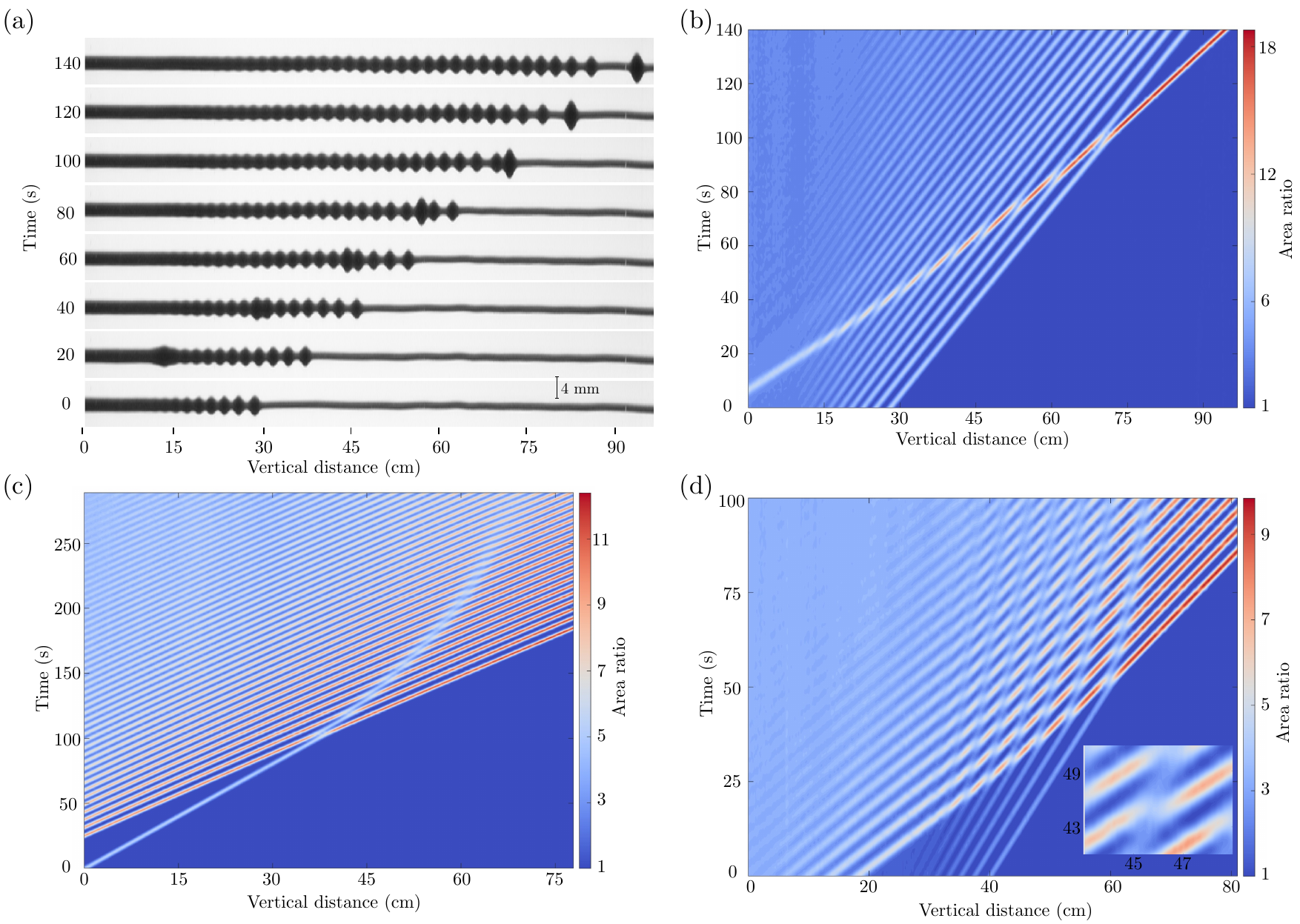}
  \caption{Interactions of solitons and DSWs.  Time-lapse images with
    aspect ratio 10:1 (a) and space-time contour (b) of DSW-soliton
    interaction revealing soliton refraction by a DSW with $a_- = 3$.
    (c) Space-time contour of the absorption of a soliton by a DSW
    with $a_- = 3.5$.  (d) DSW-DSW interaction and merger causing
    multiphase mixing (inset) and the refraction of the trailing DSW
    by the leading DSW with $a_1 = 2.5$, $a_2 = 5$.  Nominal
    experimental parameters: $\Delta \rho = 0.0971$ g/cm$^3$,
    $\mu_{\rm i} = 99.1$ cP, $\epsilon = 0.029$, $Q_0 = 0.2$ mL/min.}
  \label{fig:interactions}
\end{figure*}
The ease with which DSWs and solitons can be created in this viscous
liquid conduit system enables the investigation of novel coherent,
nonlinear wave interactions.  In Fig.~\ref{fig:interactions}, we
report soliton-DSW and DSW-DSW interactions from our conduit
experiment (see \cite{_see_????} for videos).  As in previous
experiments \cite{olson_solitary_1986,helfrich_solitary_1990}, an
isolated conduit soliton is created by the pulsed injection of fluid
on top of the steady injection that maintains the background conduit.
Figures \ref{fig:interactions}(a,b) depict the generation of a DSW
followed by a soliton.  Because solitons propagate with a nonlinear
phase velocity larger than the linear wave phase and group velocities
\cite{olson_solitary_1986}, the soliton eventually overtakes the DSW
trailing edge.  The soliton-DSW interaction results in a sequence of
phase shifts between the soliton and the crests of the modulated
wavetrain.
The soliton emerges from the interaction with a significantly
increased amplitude and decreased speed due to the smaller downstream
conduit upon which it is propagating.  The initial and final slopes of
soliton propagation in Fig.~\ref{fig:interactions}(b) demonstrate that
the soliton has been refracted by the DSW.  Meanwhile, the DSW
experiences a subtle phase shift and is otherwise unchanged.

The opposite problem of a soliton being overtaken by a DSW is
displayed in Fig.~\ref{fig:interactions}(c).  After multiple phase
shifts during interaction, the soliton is slowed down and effectively
absorbed within the interior of the DSW, while the DSW is apparently
unchanged except for a phase shift in its leading portion.  Such
behavior is consistent with the interpretation of a DSW as a modulated
wavetrain with small amplitude trailing waves that will always move
slower than a finite amplitude soliton.

Figure \ref{fig:interactions}(d) reveals the interaction of two
DSWs.  The interaction region results in a series of phase shifts due
to soliton-soliton interactions that form a quasiperiodic or two-phase
wavetrain as shown in the inset.  This nonlinear mixing eventually
subsides, leaving a single DSW representing the merger of the original
two.  The trailing DSW has effectively been refracted by the leading
DSW.

We can interpret the soliton and DSW refraction as follows.  First,
consider the overtaking interaction of two DSWs.  Denote the midstream
and upstream conduit areas $a_1 < a_2$ relative to the downstream area
$a_0 = 1$.  Equation \eqref{eq:1} implies the leading edge speeds of
the first and second DSWs are $s_1 = \sqrt{1 + 8a_1} - 1$, $s_2 = a_1
(\sqrt{9 + 8(a_2 - 1)/a_1} - 1)$.  Motivated by previous DSW
interaction studies \cite{ablowitz_soliton_2009}, we assume merger of
the two DSWs and thus obtain the leading edge speed of the merged DSW
$s_m = 4\sqrt{\frac{1}{2}(a_1+a_2)-1}-1$ connecting conduit areas
$a_0$ to $a_2$.  One can verify the interleaving property $s_1 < s_m <
s_2$, demonstrating the refraction (slowing down) of the second DSW.
If we treat the isolated soliton as the leading edge of a DSW, then we
obtain the same result for soliton-DSW refraction.

Viscous liquid conduits are a model system for the coherent dynamics
of one-dimensional superfluid-like media with microscopic-scale fluid
dynamics \cite{whitehead_wave_1988}, mesoscopic-scale solitons
\cite{helfrich_solitary_1990} and macroscopic-scale DSWs as
fundamental nonlinear excitations.  Interaction of DSWs and solitons
suggest that soliton refraction, absorption, multi-phase dynamics, and
DSW merging are general, universal features of dispersive
hydrodynamics.
The viscous liquid conduit system is a new environment in which to
investigate complex, coherent dispersive hydrodynamics that have been
inaccessible in other superfluid-like media.

\begin{acknowledgments}
  M.A.H. is grateful to Marc Spiegelman for bringing the viscous
  liquid conduit system to his attention and to Gennady El for his
  support on this work.  We thank Weiliang Sun for help with measuring
  the mass diffusion properties of fluids used in this work. This work
  was partially supported by NSF CAREER DMS-1255422 (M.A.H., D.V.A.),
  NSF GRFP (M.D.M., N.K.L.), and NSF EXTREEMS-QED DMS-1407340 (D.V.A.,
  M.E.S.).
\end{acknowledgments}

\bibliographystyle{apsrev}


\newpage

\widetext
\begin{center}
  \textbf{\large Supplemental materials for ``Observation of
    dispersive shock waves, solitons, and their interactions in
    viscous fluid conduits} \\ 
  Michelle D. Maiden, Dalton V. Anderson, Marika E. Schubert, and Mark
  A. Hoefer \\
  \textit{Department of Applied Mathematics, University of Colorado,
    Boulder CO 80309, USA} \\
  Nicholas K. Lowman \\
  \textit{Department of Mathematics, North
  Carolina State University, Raleigh, North Carolina 27695, USA}
  (Dated: \today)
\end{center}
\setcounter{equation}{0}
\setcounter{figure}{0}
\setcounter{table}{0}
\setcounter{page}{1}
\makeatletter
\renewcommand{\theequation}{S\arabic{equation}}
\renewcommand{\thefigure}{S\arabic{figure}}
\renewcommand{\bibnumfmt}[1]{[S#1]}
\renewcommand{\citenumfont}[1]{S#1}

\begin{quotation}
  In this supplemental material, background information and additional
  experimental details are provided.
\end{quotation}

\section{Background on viscous fluid conduits}
\label{sec:backgr-visc-fluid}

Principally driven by the modeling of geological and geophysical
processes, Whitehead and Luther in 1975 showed that the low Reynolds
number, buoyant dynamics of two fluids with differing densities and
viscosities could lead to the formation of stable fluid filled pipes
or conduits of upwelling fluid \cite{whitehead_dynamics_1975}.  In
1984, McKenzie derived a system of equations describing the dynamics
of melt (magma) within a deformable matrix (rock) in the upper Earth's
mantle \cite{mckenzie_generation_1984}.  These equations treat the
magma dynamics as the flow of a low Reynolds number, incompressible
fluid through a more viscous, permeable matrix that is modeled as a
compressible fluid due to compaction and distension.  There are two
model parameters $(n,m)$ resulting from constitutive power laws that
relate the porosity to the matrix permeability and viscosity,
respectively. Soon after, it was realized that the asymptotically
reduced, one-dimensional McKenzie equations, or magma equation,
exhibits solitary wave solutions \cite{scott_magma_1984}.  A
connection between the laboratory fluid systems explored by Whitehead
and Luther and the magma equation was realized in
\cite{olson_solitary_1986a,scott_observations_1986a} where the conduit
equation studied in the present work (eq.~1 in the primary manuscript)
was derived from physical arguments, the soliton amplitude-speed
relation was verified experimentally, and the approximately elastic
solitonic interaction property was observed.  The conduit equation
corresponds exactly to the magma equation when $(n,m) = (2,1)$.  Since
that time, there have been a number of experimental and theoretical
works on viscous fluid conduits, principally focused upon the dynamics
of solitons (see, e.g.,
\cite{whitehead_magma_1990,simpson_asymptotic_2008} and references
therein).  It is now known that the one- and two-dimensional solitary
wave solutions to McKenzie's equations are unstable to transverse
perturbations, leading to the formation of fully three-dimensional
solitary waves \cite{wiggins_magma_1995}.

In this work, rather than emphasize the connection to McKenzie's
equations and magma dynamics, we consider the dynamics of viscous
fluid conduits as a model \textit{dispersive hydrodynamic system}
where hydrodynamic nonlinearity is balanced by dispersive effects.
Such systems are plentiful in the natural world, as commented upon in
the introduction of the accompanying manuscript.  In addition to
solitons, dispersive shock waves (DSWs) are fundamental nonlinear
excitations in dispersive hydrodynamic media (see the review
\cite{el_dispersive_2016}).  The first numerical studies of DSWs in
the McKenzie equations was undertaken by Spiegelman
\cite{spiegelman_flow_1993b}.  Whitham modulation theory was later
used to describe DSWs in the small \cite{elperin_nondissipative_1994}
and large \cite{marchant_approximate_2005,lowman_interactions_2014a}
amplitude regimes of the magma and, in particular, the conduit
equation.  The present work represents the first experimental study of
conduit DSWs.

\section{Poiseuille flow relation}
\label{sec:pois-flow-relat}

\begin{figure}
  \centering
  \includegraphics{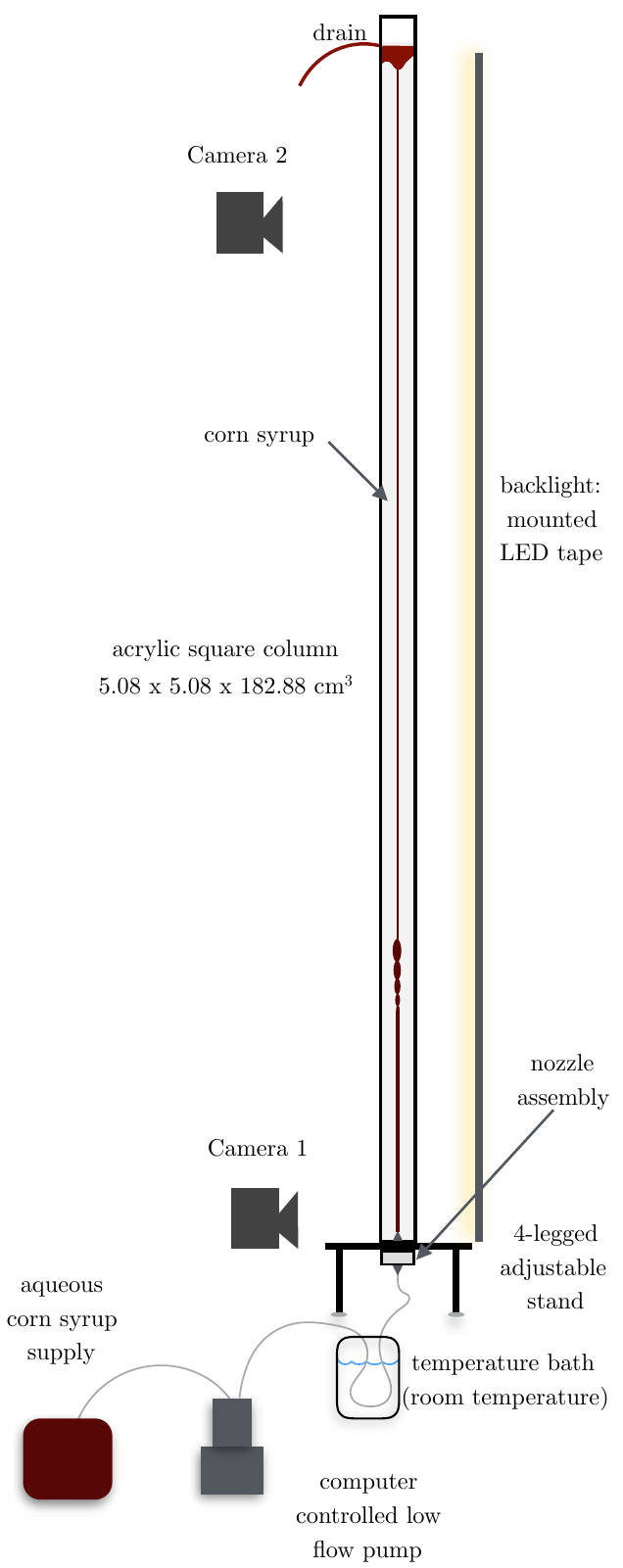}
  \caption{Schematic of the conduit experimental apparatus.}
  \label{fig:schematic}
\end{figure}

The conduit experimental data are obtained by injecting through a 0.22
cm inner diameter nozzle an approximately 7:2:1 mixture of corn syrup
(Karo brand light), water, and black food coloring (Regal brand) into
the bottom of a 2 m tall acrylic, 25.8 cm$^2$ square column filled
with corn syrup (3:2 mixture of Golden Barrel brand 42 dextrose
equivalent and Karo brand light for data of Figure 2, pure Karo brand
light for Figures 1, 3, 4).  The fluid temperature near the top of the
fluid column was measured to be $22.2 \pm 0.7$ deg C across all
experimental trials.  A computer controlled piston pump (Global FIA
milliGat LF pump with MicroLynx controller) was used to inject fluid
through a room temperature water bath followed by the nozzle.  See
Fig.~\ref{fig:schematic} for an experimental schematic.  When the
injected fluid reaches the top of the fluid column, it pools on top of
the external fluid and very slowly begins to diffuse downward.  We
periodically removed the pooling fluid with a syringe.  Steady
injection results in a vertically uniform liquid filled pipe or
conduit conforming to Poiseuille flow
\cite{whitehead_dynamics_1975,scott_observations_1986a}.  We allowed
the conduit to stabilize (straighten) by steady injection over a
period of 36 hours for the data in Fig.~2 and 15 hours
for the other data.

\begin{figure}
  \centering
  \includegraphics{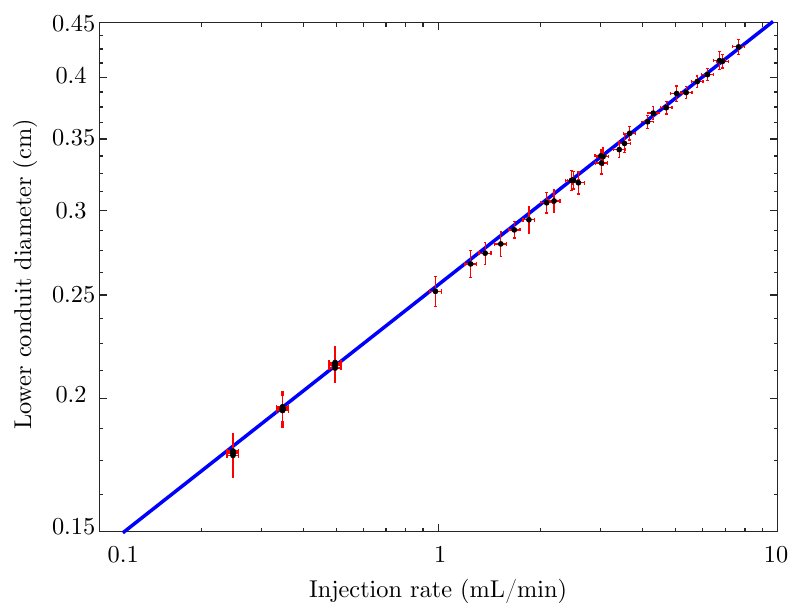}
  \caption{Demonstration of Poiseuille flow in a steady viscous fluid
    conduit.  Log-log plot of measured conduit diameter $D$ near
    injection site versus volumetric flow rate $Q$ (dots) and the
    relation $D = \alpha Q^{1/4}$ with the measured value $\alpha =
    0.2557$ (cm$\cdot$min)$^{1/4}$ (solid) corresponding to $\mu_{\rm i}
    = 80.4$ cP, $\Delta \rho = 0.1305$ g/cm$^3$.  A least squares fit
    gives $\alpha = 0.2548$ (cm$\cdot$min)$^{1/4}$, which translates
    to the fitted viscosity $\mu_{\rm i} = 79.0$ cP, within the 2\%
    error tolerance of our rotational viscometer.}
  \label{fig:poiseuille}
\end{figure}

\begin{figure}
  \centering
  \includegraphics{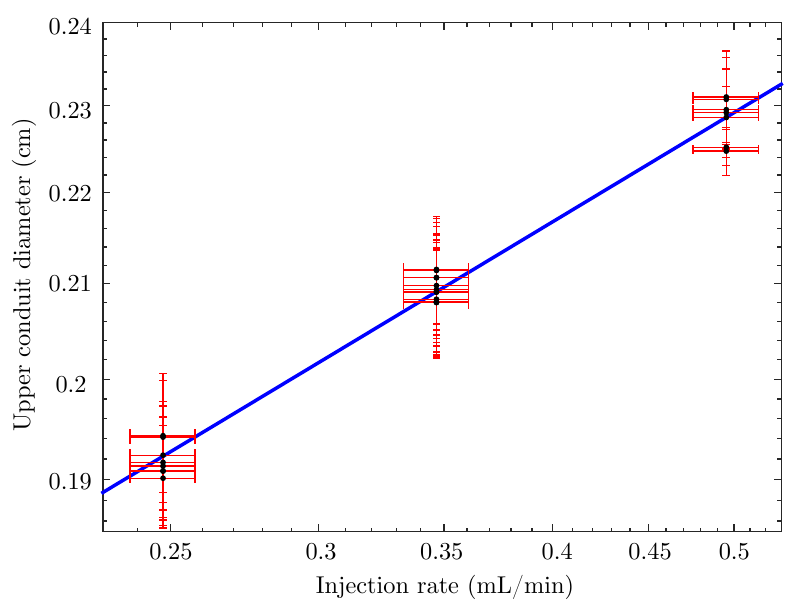}
  \caption{Poiseuille flow fit approximately 120 cm up the fluid
    column.  Downstream conduit diameters $D$ extracted from digital
    images (dots) and a least squares fit to the Poiseuille flow
    relation $D = \alpha Q^{1/4}$ with $\alpha = 0.2688$
    (cm$\cdot$min)$^{1/4}$ (solid). The fit corresponds to the
    interior viscosity $\mu_{\rm i} = 104$ cP, an increase from its
    measured value $\mu_{\rm i} = 80.4$ cP. This can be explained by the
    shear thinning properties of corn syrup as described in Methods.}
  \label{fig:poiseuille_high}
\end{figure}

\begin{figure}
  \centering
  \includegraphics{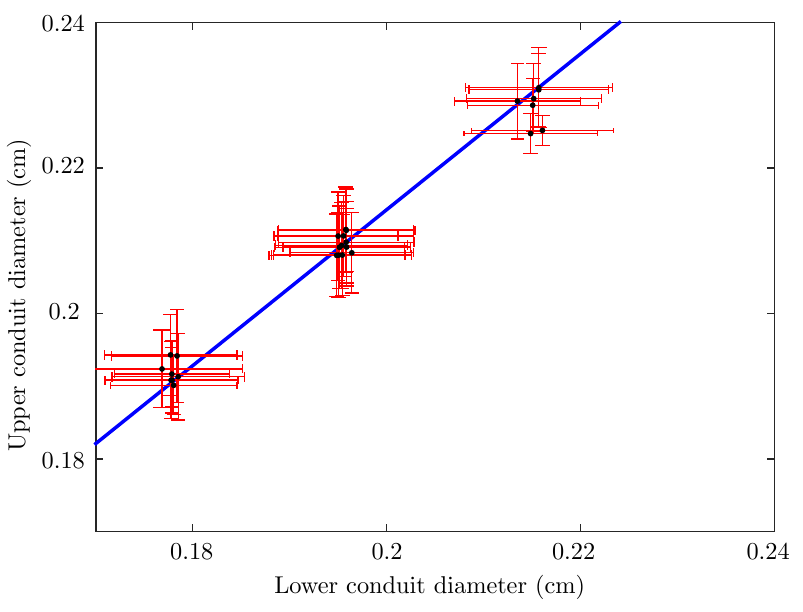}
  \caption{Comparison of conduit diameter at different locations along
    the fluid column.  Measurements (dots) and the linear fit $D_{\rm
      top} = m D_{\rm bottom}$ (solid) with $m = 1.07$ corresponding
    to a 7\% increase in the conduit diameter.  The lower (upper)
    diameter was measured approximately 6 cm (120 cm) above the
    injection site.}
  \label{fig:bottom_top_diam}
\end{figure}

The quantitative data in Fig.~2 exhibits typical conduit
diameters of one to four millimetres and Reynolds numbers in the range
Re $= \rho_{\rm i} U_0 L_0/\mu_{\rm i} \in (0.06,2.6)$, where
$\rho_{\rm i}$ is the intrusive fluid density.  We can set the conduit
diameter $D$ via the volumetric flow rate $Q$ according to a
Hagan-Poiseuille relation \cite{whitehead_dynamics_1975} $D = \alpha
Q^{1/4} = (2^7 \mu_{\rm i} Q)^{1/4}/(\pi g \Delta\rho)^{1/4}$.
Digital images of the conduit are processed to extract the conduit
diameter.  The conduit edges are determined from local extrema of the
differentiated grayscale intensity image using centred differences in
the direction normal to the conduit interface.  We confirm the
Poiseuille flow relation $D = \alpha Q^{1/4}$ for the trials of
Fig.~2 approximately 6 cm above the fluid injection site
with no fitting parameters (Fig.~\ref{fig:poiseuille}).  In
Fig.~\ref{fig:poiseuille_high}, we show the fit of the Poiseuille flow
relation to the same conduit, imaged approximately 120 cm above the
injection site.  The difference between the externally measured
viscosity $\mu_{\rm i} = 80.4$ cP and the value $\mu_{\rm i} = 104$ cP
from a fit to the Poiseuille flow relation can be explained by the
non-Newtonian, thixotropic (shear thinning) properties of corn syrup.
At the injection site, the diluted corn syrup experiences heightened
shearing, similar to our rotational viscometer measurements
(Brookfield DV-I prime viscometer).  Further up the fluid column,
there is less shearing so the fluid increases in viscosity and leads
to a dilation of the conduit.  The conduit consistently has a measured
diameter in the upper fluid column that is 7\% larger than its value
near the injection site as shown in Fig.~\ref{fig:bottom_top_diam}.
The results in Fig.~2 of the main text use the measured
value of $\Delta \rho$ and the fitted value $\mu_{\rm i} = 104$ cP.

\section{DSW and soliton injection protocol.}
\label{sec:dsw-inject-prot}

By adiabatically changing $Q$, we introduce perturbations to the
conduit that subsequently propagate along the interface, allowing for
the generation of conduit solitons
\cite{olson_solitary_1986a,whitehead_wave_1988a,helfrich_solitary_1990a,lowman_interactions_2014a}
and DSWs.  The injection rate profile for solitons is generated by
computing a conduit solitary wave solution $a_{\rm soliton}(z-ct-z_0)$
with speed $c$ and initial center $z_0$ to eq.~(1).  This profile is
converted to the dimensional diameter $D_{\rm soliton} = 2\sqrt{a_{\rm
    soliton}A_0/\pi}$ and then the volumetric flow rate profile
$Q_{\rm soliton} = (D_{\rm soliton}/\alpha)^4$, evaluated at the
injection site.

\begin{figure}
  \centering
  \includegraphics{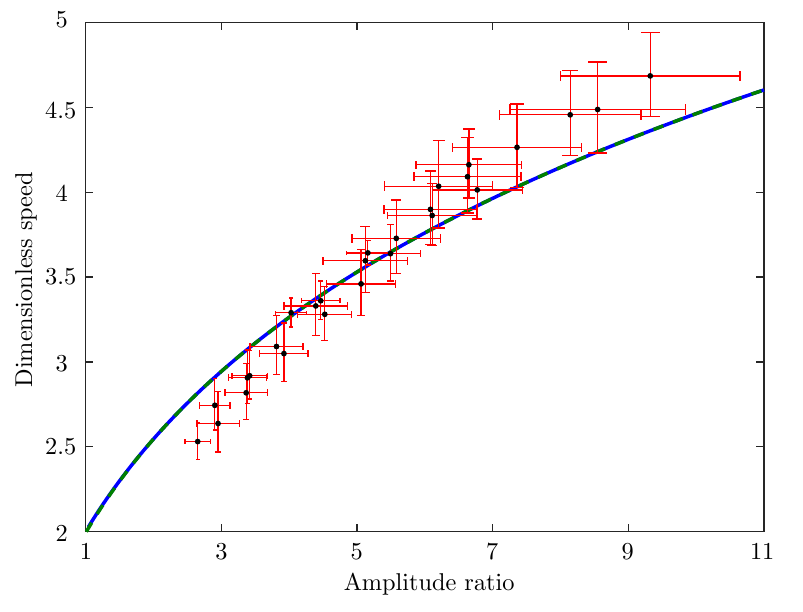}
  \caption{DSW leading edge speed versus amplitude.  The observed
    values (dots) and the theoretical soliton dispersion relation
    (solid) corresponding to Fig.~2(a,b) in the main
    text.  The deviation at large amplitudes is consistent with
    previous studies of isolated solitons \cite{olson_solitary_1986a}.}
  \label{fig:ampl_speed}
\end{figure}
The volumetric flow rate profile $Q_{\rm DSW}$ that we use to create
DSWs is determined as follows.  We initialize the dispersionless
conduit equation $a_t + \left (a^2 \right)_z = 0$ with a step in
conduit area from $a_-$ to 1, left to right, at a desired distance
from the nozzle $z = z_b$.  Evolving this initial value problem
backward in time results in a non-centered rarefaction wave that can
be related to the volumetric flow rate profile via
\begin{equation*}
  Q_{\rm DSW}(\tau) = Q_0
  \begin{cases}
    1 & t \le 0 \\
    (1 - 2 \tau U_0/Z_b)^{-2} & 0 < \tau U_0/Z_b < (a_- - 1)/2 a_- \\
    a_-^2 & \mathrm{else}
  \end{cases},
\end{equation*}
where $Q_0$ is the downstream flow rate, $\tau$ is the dimensional
time, and $Z_b = L_0 z_b$ is the dimensional breaking distance from
the injection site.  We find that this provides adequate control over
the breaking location.

Each DSW trial in Fig.~2 was initiated after a sufficient
waiting period, typically 5 minutes, to allow the previous trial's
conduit diameter to stabilize to the expected steady value.  The
downstream flow rates utilized for the data in Fig.~2
were nominally $Q_0 \in \{0.25,0.35,0.5 \}$ mL/min.  Three digital SLR
cameras were utilized, two Canon EOS 70D camers outfitted with Tamron
macro lenses positioned just above the injection site (camera one) and
at approximately 120 cm above the injection site (camera two).  The
third camera (Canon EOS Rebel T5i), outfitted with a zoom lens, was
used to image the entire vertical length of 120 cm from the injection
site.  The fluid column was backlit with strip LED lights behind LEE
LE251R white diffusion filter paper.  Each DSW trial was initiated
with an image of the conduit at the injection site followed by the
injection protocol $Q_{\rm DSW}$.  The third camera was then set to
image the full column every second throughout the trial. Just after
the injection protocol reached the maximum rate $a_- Q_0$, an image of
the conduit from camera 1 was taken.  Just prior to the arrival of the
DSW leading edge within the viewing area of camera two, an image of
the downstream conduit was taken, followed by a dozen or more images
taken in rapid succession of the DSW leading edge.

\section{Determination of DSW speed and amplitude.}
\label{sec:exper-param-prot}

The leading edge of the DSW amplitude, normalised to the downstream
area, is determined from the digital images of camera 2 without
appealing to any fluid parameters.  We compute the conduit edges as
for the steady case, using extrema of the differentiated image
intensity normal to the conduit interface.  Some image and edge
smoothing is performed to remove pixel noise.  The number of pixels
across the diameter of the leading edge DSW peak is calculated and
normalized by the diameter of the downstream conduit.  Squaring this
quantity gives the leading edge DSW amplitude shown in Fig.~2.  We
calculated the leading edge DSW speed from the images of camera three
toward the end of the trial.  We nondimensionalise the speed by the
characteristic speed $U_0 = L_0/T_0 = g A_0 \Delta \rho/(8\pi
\mu_{\rm i})$, where we use the measured values of the downstream flow
area $A_0$ from camera two and $\Delta \rho$.  The fitted value for
$\mu_{\rm i}$ is used, as described in the earlier section on Poiseuille
flow.

\section{Mass diffusion.}
\label{sec:mass-versus-momentum}

The injected and external fluids are miscible so there is unavoidable
mass diffusion across an interface between the two.  Using a
procedure similar to that described in \cite{ray_determination_2007},
we estimate the diffusion constant $\tilde{D}$ between a 7:3 corn
syrup, water mixture and pure corn syrup (Karo brand light) to be
approximately $1.2 \times 10^{-6}$ cm$^2$/s.  Combining this with
typical flow parameters, we estimate the P\'{e}clet and Schmidt
numbers for the trials of Fig.~2 to lie in the range Pe = $L_0
U_0/\tilde{D} \in (2.1 \times 10^{4},7.9 \times 10^5)$ and Sc = Pe/Re
$\approx 5.2\times 10^5$.  The advective time scale for Fig.~2 trials
is in the range $T_0 \in (1.6,5.6)$ s.  We therefore estimate that
mass diffusion begins to play a role after approximately 9 hours,
whereas the time scale of an experimental trial is less than 10
minutes.

\bibliographystyle{apsrev}


\end{document}